\documentclass{ws-acs}
\usepackage{url}
\begin{document}

\makeatletter
\def\@biblabel#1{[#1]}
\makeatother

\markboth{A. Guazzini, D. Vilone, F. Bagnoli, T. Carletti, R. Lauro Grotto}{Cognitive network structure: an experimental study}

%
\catchline{}{}{}{}{}
%

\title{Cognitive network structure: an experimental study}

\author{\footnotesize Andrea Guazzini}

\address{Department of Psychology, Universit\`a degli studi di Firenze, and
Istituto di Informatica e Telematica (IIT), Consiglio
Nazionale delle Ricerche (CNR), Via di San Salvi 12, 50135, Firenze, Italy.\\
andrea.guazzini@unifi.it}

\author{Daniele Vilone}

\address{Institut de F\'isica Interdisciplin\`aria i Sistemes Complexos, Campus Universitat de les Illes Balears, E-07122 Palma de Mallorca, Spain.\\
daniele.vilone@gmail.com}

\author{Franco Bagnoli}

\address{Dipartimento Energetica and CSDC, Universit\`a degli studi di Firenze, via S. Marta 3, I-50139 Firenze, Italy. Also INFN sez. Firenze,\\
franco.bagnoli@unifi.it }

\author{Timoteo Carletti}

\address{Namur Center for Complex Systems, Facult\'es Universitaires Notre-Dame de la Paix, 8 rempart de la vierge, B5000 Namur, Belgium.\\
timoteo.carletti@fundp.ac.be}

\author{Rosapia Lauro Grotto}

\address{Department of Psychology, Universit\`a degli studi di Firenze,  Via di San Salvi 12, 50135, Firenze, Italy.\\
grotto@psico.unifi.it}

\maketitle

\begin{history}
\end{history}

\begin{abstract}
In this paper we present first experimental results about a small group of people exchanging private and public messages in a virtual community. Our goal is the study of the cognitive network that emerges during a chat seance. We used the Derrida coefficient and the triangle structure under the working assumption that moods and perceived mutual affinity can produce results complementary to a full semantic analysis. The most outstanding outcome is the difference between the network obtained considering publicly exchanged messages and the one considering only privately exchanged messages: in the former case, the network is very homogeneous, in the sense that each individual interacts in the same way with all the participants, whilst in the latter the interactions among different agents are very heterogeneous, and are based on \lq\lq the enemy of my enemy is my friend\rq\rq strategy.
Finally a recent characterization of the triangular cliques has been considered in order to describe the intimate structure of the network. Experimental results confirm recent theoretical studies indicating that certain 3-vertex structures can be used as indicators for the network aging and some relevant dynamical features.
\end{abstract}


\section{Introduction}

The behavior of a single human being is not always reducible to the physiological and psychological processes alone. The presence of a large number of factors affecting the psychological processes has afflicted the research  of both the informative experimental observables and the simple and effective models of cognition. The factorial analysis and other sophisticated statistical procedures \cite{Cattell1966,Comrey1992} clearly demonstrated the complexity of the human behavior and the necessity to include in such analysis the role of the environmental information factors, elaborated and filtered by the cognition. 

While in classical cognitive models the influence of the individual characteristics on the social dynamics was considered to be predominant, the understanding of the role of the group, drove the Social Cognition Psychology toward the definition of new concepts \cite{Newman,Fortunato,Zhang1994}. Since the works of Kurt Lewin~\cite{Lewin1943}, the \textit{Psychological Field}, hereafter PF for short, has been the main concept that has merged the cognitive psychological theories -- \textit{i.e. the study of the single human mental processes} and the social psychology constructs. 

At the same time, sociology and social psychology offered a new perspective: the group structure of interacting humans is a crucial factor affecting the individual psychological processes and the efficiency of problem solving tasks~\cite{Leavitt1951,Bavelas1950}; moreover this provide also a reliable proxy the the group dynamics~\cite{Bion,Vilone2011}.

The PF theory essentially underlines that  individual's acts, for instance thoughts, decisions, behaviors, are originated by a series of \lq\lq fast, frugal and smart\rq\rq mental processes named \textit{cognitive heuristics}~\cite{Gigerenzer1996,Gigerenzer2011,Simon1976} that are affected by the individual internal characteristics but also by the \lq\lq psychological environmental features\rq\rq. The latter being defined by the \textit{social cognition}~\cite{Neisser1967,Festinger1950} as the ensemble of all external factors affecting the human decisional processes. The experimental measures of PF is a complex task, given human ability in detecting non-semantic messages and their influence on the emotional content. However, nowadays there is the possibility of performing experiments using a computer interface, and therefore to measure many components of communication, the only source of ambiguity remaining in the semantic contents of the exchanged messages.

The goal of our experiment was to discover the strategies (heuristics) that the subjects adopted to cope with the social problem posed by the tasks: assessing their own affinity with others in presence and absence of an imposed topic. In the second task we furthermore asked the participants to reach consensus at least in some groups.
In particular, we wanted to design an experimental controlled setting and a set of related procedures, in order to detect the effects due to the subjects’ PF features. Moreover we made the working assumption that moods and perceived mutual affinity can produce results complementary to a full semantic analysis. Our results will rely on the perception of the PF by the involved agents. We have thus developed a chatline allowing us to record the interactions among a group of ten participants~\cite{Guazzini2010}. The environment mimics an unstructured group of initially unknown agents. Each member can communicate with all the agents at once, \textit{i.e.} through a public communication channel; a second private channel exists, allowing each agent to choose to which address his/her messages.
The chat interface has been designed in order to easily detect the mood and the target of the messages, but also the \lq\lq perception of the others\rq\rq , through the exploitation of dedicated tools on the interface (i.e. the private radar). Eventually all such information can be merged together to spatially represent the perception of the social geometry of the group, that we consider to be affected mainly by the subjects’ PF.

We also propose an analysis of the group stability by studying the excess/defect of polarized triangles in the resulting social network, i.e. triangles whose links can have positive and/or negative content. Because of our working assumption, the social network has been built considering both the number of messages and their mood. Finally the number of the three vertex cliques (i.e. triangles) can be easily carried out by standard techniques. Szell et al. \cite{Szell2011} have shown that such structures can be used to accurately describe the cognitive network and can be related to the cognitive task which is shaping the network structure itself. Some of these structures can became \lq\lq unstable\rq\rq  depending on the constraints of the task, and this should happen when they are not effective or useful to solve the social problem the agents are faced to. Hence the excess/defect of such polarized triangles constitutes a good experimental observable for the characterization of the cognitive tasks and of the cognitive strategies used by both the subjects and the \lq\lq group entity\rq\rq to solve the social problem~\cite{Brautbar2011}.

The paper is organized as follows. In Section \ref{Section: Experimental_framework} we present the experimental tool and we define the main observables exploited in the data analysis. The samples and the experimental procedures are then introduced and the statistical and network analyses are described in subsections \ref{Sample and Procedures} and \ref{Statistical and Structural Analysis}. Finally,  the results are illustrated and discussed in sections \ref{Section: Results} and \ref{Section: Discussion}.

\section{The Experimental framework}
\label{Section: Experimental_framework}

The proposed framework is composed by a chatline interface, a computer room and a server used to record all the activities: messages exchanged, associated mood, personal perception of mutual affinities, etc ...

\subsection{Chatline}
\begin{figure}
\centering
\includegraphics[width=10cm]{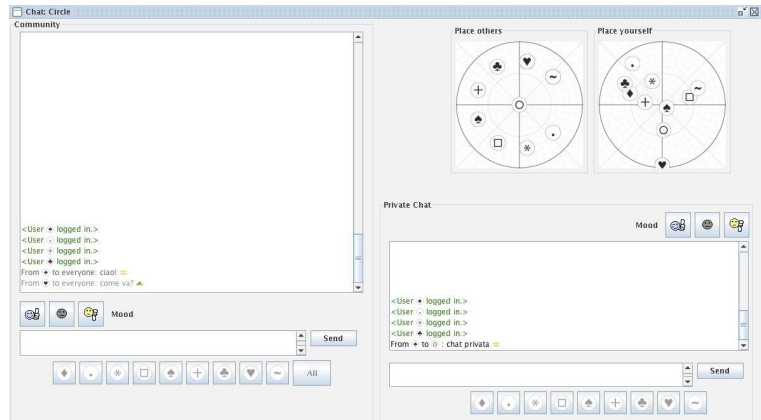}
\caption{Chat interface}\label{fig:Interfaccia}
\end{figure}

The chatline interface is composed by two textual windows, one for communicating with the rest of participants in a public way, and one to communicate with a selected one in a private way (Fig. \ref{fig:Interfaccia}).

To keep agent's anonymity and to remove the influences due to the visual perception of each agents, we included in the interface two \lq\lq radars\rq\rq , where the avatars representing the other participants may be placed. Radars are labelled as \lq\lq public\rq\rq and \lq\lq private\rq\rq: the public one can be seen by all other agents and each agent can only move his/her own avatar. The agents have to move the avatars according their perceived proximity with the others.
On the other hand, on each private radar, owner agent can move all other avatars while his/her own is blocked in the center. In this way we offer an equivalent of the non-verbal communications (the public radars) similar to change place so to be closer to a given person, and a mnemonic aid (the private radar) for the representation of others' identities and their perceived social proximity, as seen by each user. 

In order to reinforce this interpretation, messages are darker the closer is the sender to receiver in the public radar, and vice versa becoming clear the farther the sender. This is because as in real life a person often tend to interact more with people which feels closer to him/her, then here we wanted to have the same effect. Moreover, this setting makes the communication among individuals more efficient.
Finally, as already states, each message can be characterized by the sender with a \lq\lq mood\rq\rq, represented by a small icon with thumb up, down or neutral. For an exhaustive description of the tool the interested reader can consult reference~\cite{Guazzini2010}.

From each communication channel, private and public textual chats and private radar, we have extracted the relevant events from the log file, and computed several indicators to describe both the communication and the cognitive network. Since the semantic content of a message crucially depends on the cultural context, and having already scheduled future experiments with people coming from different countries (and cultures), we decided to base our analysis only on the timing and number of messages exchanged by agents~\cite{Nguyen2009}. Of course the semantic aspects represent a very relevant dimension to analyze and such analysis will require a dedicated work.

\subsection{Sample and Procedures}
\label{Sample and Procedures}

The experimental sample is composed by $100$ volunteer students of the Psychology Faculty of the University of Florence (Italy). All participants were  Italians with an age ranging from $20$ to $26$.  

These $100$ participants were randomly assigned  to one of the $10$ experimental sessions each one composed thus by $10$ individuals; we stratified the sample in equivalent clusters in order to obtain sessions where people do not know each other. The sample was also stratified on the base of the gender to obtain experiments with $5$ males and $5$ females. The experimental setting was located into a computer room of the Psychology Faculty, and the ten computer locations  have been enclosed in kiosks so that no subject could see the others. The experiment duration was of $60$ minutes organized in two parts, a first $15$ minutes long training phase where the software interface was illustrated and the task requirements administered, and a $45$ minutes long experimental phase of virtual interaction.

After a random assignment of a nick name and of a position in the laboratory, all the subjects were instructed throughout a standardized slide presentation on how to use the chat interface and  the task requirements. They were asked to accomplish one of the two tasks during the session, labelled respectively as Blank and Topic modality.

The Blank modality was designed as \lq\lq control task\lq\lq and consequently its target was to introduce the smallest possible number of constraints and possible biases. Accordingly, we selected a daily solved social problem, estimating the affinity with another subject by freely chatting for $45$ minutes. The participants could interact using  public and private rooms, and were explicitly asked to assess  their perceived affinity with the others and reporting them on the private radar before the end of the experiments. The affinity with someone was defined as the perceived degree of similarity in terms of opinions, beliefs and attitudes.

The Topic modality was designed to introduce a first constraint affecting the same task as before (and maybe/presumably the psychological field).  Subjects were asked to participate in a role game where they belonged to an \textit{ethic committee} that was charged to reform the law that controls the researches involving animals. The requirements were  to discuss about the given topic, developing before the end of the experiment one or more shared ethical positions, and assessing the affinity space accordingly~\cite{Crowne}.

\subsection{Statistical and Structural Analysis}
\label{Statistical and Structural Analysis}

By analyzing the log files of the experimental sessions we extracted all the observables useful to describe the communications and the perceived affinity among subjects.

We define a  Markov matrix where each element $M^t_{ij}$ represents the probability  to observe a message from the subject $i$ to the subject $j$ at time $t$,
\begin{equation}
\label{eq:Message_Matrix}
M_{ij}^{t}=\frac{W_{ij}^{t}}{\sum_{k=1}^{N} W_{ik}^{t}}\, ,
\end{equation}
where $W_{ij}^t$ is the number of messages sent by $i$ to $j$ until time $t$. 

The major role in our investigation is played by the affinity space, that is the $n-1$ dimensional space determined by the overlapping of the $n$ subjects' affinity spaces. Since subjects were asked to indicate their perceived affinity with the others by using their private radars, affinity is assumed to correlate with the distance among the icons. At the beginning, the distance among icons for all the private radars is  $0.25$, half of the radius of the radar circle. We define the affinity variable as
\begin{equation}
\label{eq:Affinity_distance}
\delta_{ij}^{t}=2d_{ij}^t,
\end{equation}
where $d_{ij}^t \in (0,0.5)$ is the distance between individual $i$ and $j$ in the private radar of individual $i$ and $\delta_{ij}^t \in (0,1)$ is the perceived affinity of $j$ that $i$ has at time $t$. For sake of simplicity we  consider only the final state of the affinity space, that is, the final state of the ten private radars at the end of the $45'$ experiment. 

The social distance $\delta_{ij}^{t}$, Eq. \eqref{eq:Affinity_distance}, has been used to represent the affinity network perceived by the subjects. Finally, $\delta_{ij}$ has been used to study the structure of the cognitive networks shaped by the tasks and the communicative precursors of the final perceived affinity. 

The structural analysis has considered the networks defined by the spaces taken into account. 

For all the subjects the normalized average activity for the $9$ spaces were calculated. Concerning the private radar, since each subject has his/her own strategy for managing his/her private radar, \textit{i.e.}, imposes a personal metrics to the displacements of icons, we first normalized all the rough original distances, 
\begin{equation}
	\label{eq:Affinity_normalization}
	\widehat{\delta_{ij}}=\frac{\delta_{ij}-\min_k(\delta_{ik})}{\max_k(\delta_{ik})-\min_k(\delta_{ik})},
\end{equation}
and converted them to binary values,  
\begin{equation}
	\label{eq:Affinity_recoding}
	\Delta_{ij}=\begin{cases} 
	  +1, & \mbox{if }\widehat{\delta_{ij}} < 0.5,\\
	   -1, & \mbox{if }\widehat{\delta_{ij}} > 0.5.
	\end{cases}
\end{equation}

At end of this procedure we have two groups for each individual, according to the value  of $\Delta_{ij}$. This variable was used to investigate the mutual affinity of the subjects, comparing the groups defined from this observables with respect to the registered communication dimensions. We adopted the Student \textit{t}-test for independent samples in order to compare the averages of the two groups, considering only those dimensions characterized by a $t$-value larger than $2$. 

The group level analysis required a different approach due to the peculiar setting represented by the small group dynamics. The most used indicators for describing the group structure are  based on the concept of cluster, but in psychology it is known that there are different nested structures, with different strengths, sizes and characteristics for the human social networks. In order to take into account this fact we defined a clustering spectrum as follows. Given a threshold parameter $p$ in the interval $(0,1)$, the number of final clusters for the space considered is obtained by comparing the elements of matrix $M$ to $p$.
\begin{equation}
\label{pippo}
  \hat{M}_{ij} = \begin{cases}
    1 & \text{if $M_{ij}^{t_{fin}} > p$,}\\
    0 & \text{if $M_{ij}^{t_{fin}} < p$.}
    \end{cases}
\end{equation}
In practice, we define two individuals as connected if when the experiment has finished their probability to exchange a message is larger then a threshold $p$.
The group structure can be consequently represented by plotting the number of clusters in $\hat{M}$ versus $p$.

Given the small size of the analyzed networks, we considered not only the number of clusters but also their relative size by  computing the Derrida coefficient~\cite{Wuensche:2011},
\begin{equation}
\label{eq:Derrida_coeff}
D_{\hat{M}}(p)=\sum_{i=0}^k \frac{S_{i}^2(p)}{N^2},
\end{equation}
where $D_M(p)$ is the Derrida coefficient for the matrix $M$ with threshold $p$, $S_i(p)$ is the size of the cluster $i$ and $N$ is the size of entire network.

To obtain a more in-depth representation of the cognitive network, we develop a criterion based on the affinity space to characterize the links among subjects as positive, neutral or negative. First we computed for each subject the average distance of the icons from the centre in his/her private radar,
\begin{equation}
\label{eq:Average_Distance}
\overline{\delta_i}=\frac{\sum_{j\neq i;j=1}^N \delta_{ij}}{N-1},
\end{equation}
so  to define an individual radar metrics.  Then we considered as positive the relations with those icons positioned closer than the average distance. The relations were supposed as mild negative if the distances between icons were larger than the averages but lower than the icons initial distance $(0.25)$, while they were labelled  negative if the icons were also farther away from the center than the initial distance. Finally we built the resulting network adjacency matrix $A_{ij}$ with the following rules: we create a positive link between two nodes if both the relations were positive (i.e. $\delta_{ij} < \overline{\delta_i}$ and $\delta_{ji} < \overline{\delta_i}$), in all the remaining cases we create a negative link.

\begin{figure}[t]
\centering 
\begin{minipage}[c]{.45\textwidth}
	\centering
	\includegraphics[width=0.5\textwidth]{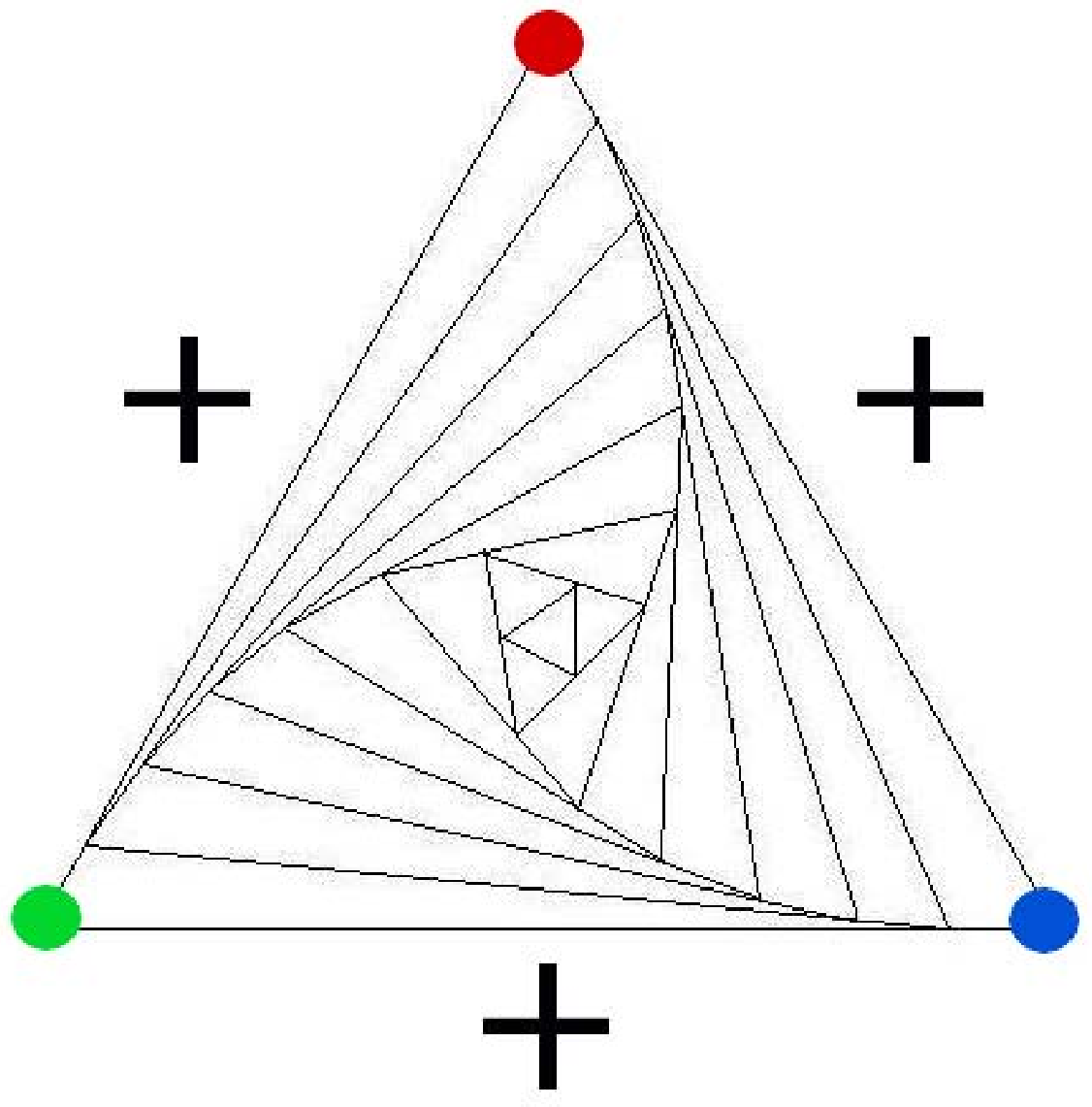}
	\vspace{-5mm}
\end{minipage}%
\hspace{10mm}%
\begin{minipage}[c]{.45\textwidth}
	\centering
 	\includegraphics[width=0.5\textwidth]{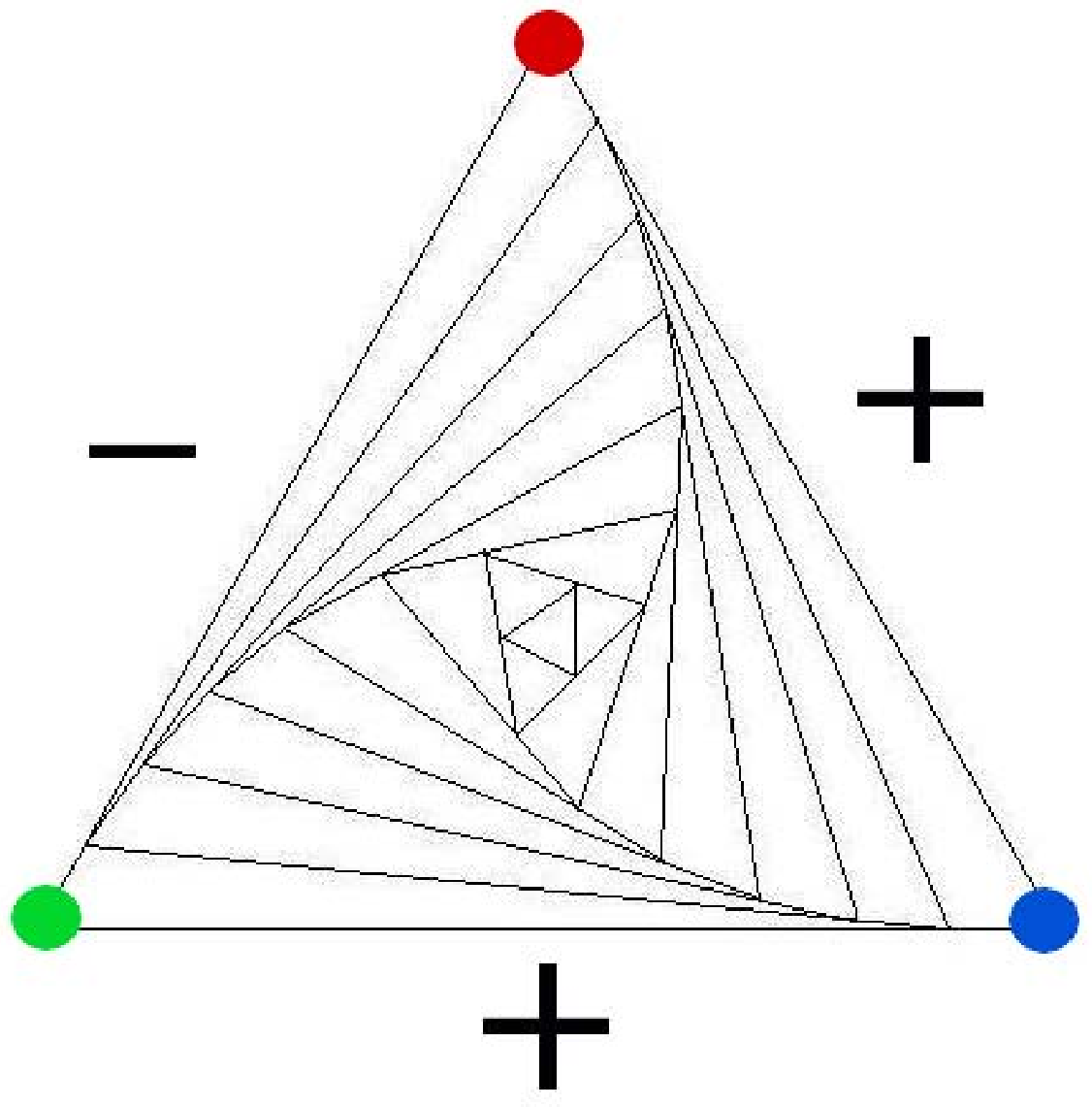} 
 	\vspace{-5mm}
\end{minipage} 
\vspace{10mm}
\hspace{10mm}%
\begin{minipage}[c]{.45\textwidth}
	\centering
	\includegraphics[width=0.5\textwidth]{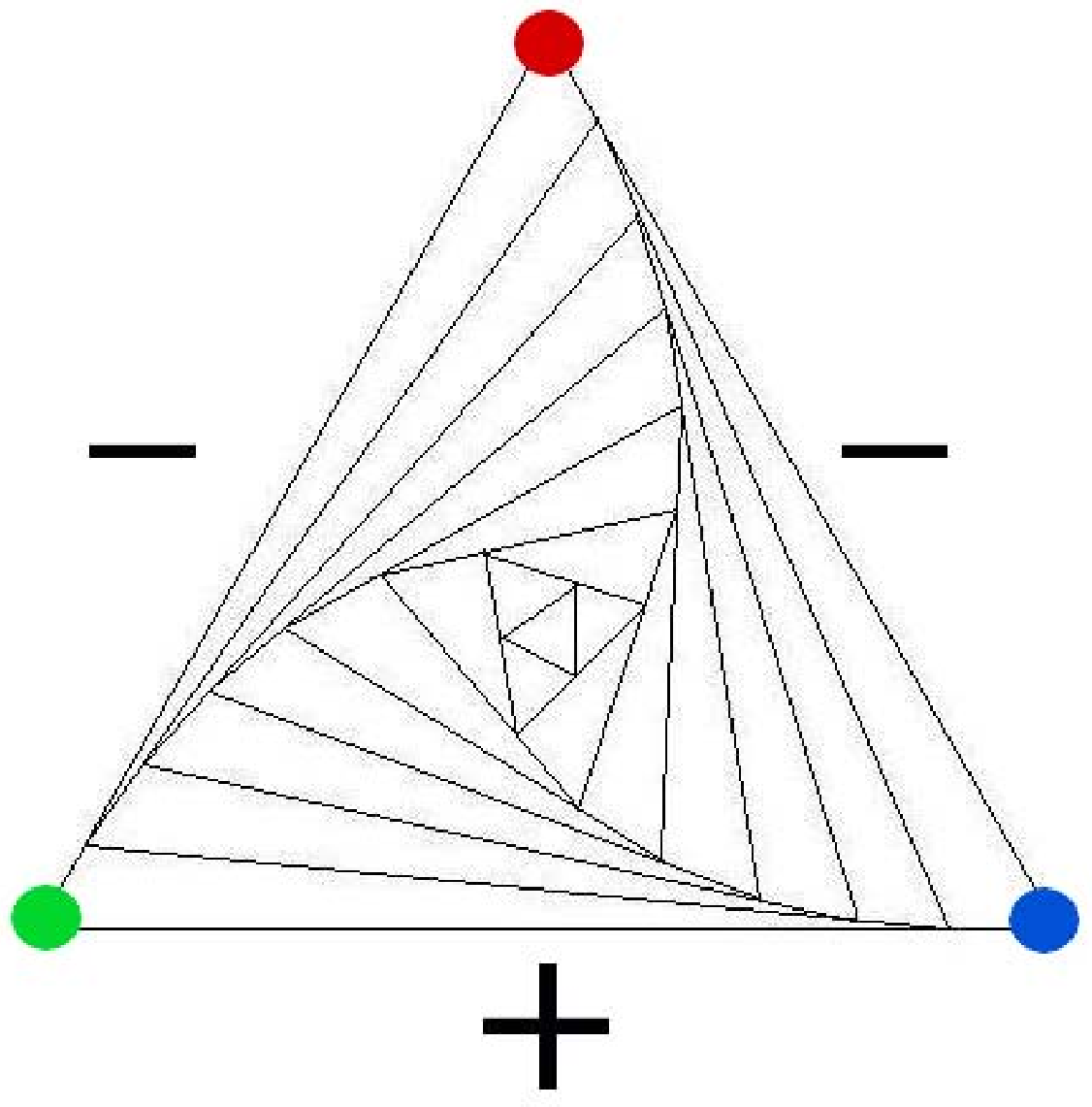}
	\vspace{-5mm}
\end{minipage}%
\hspace{10mm}%
\begin{minipage}[c]{.45\textwidth}
	\centering
 	\includegraphics[width=0.5\textwidth]{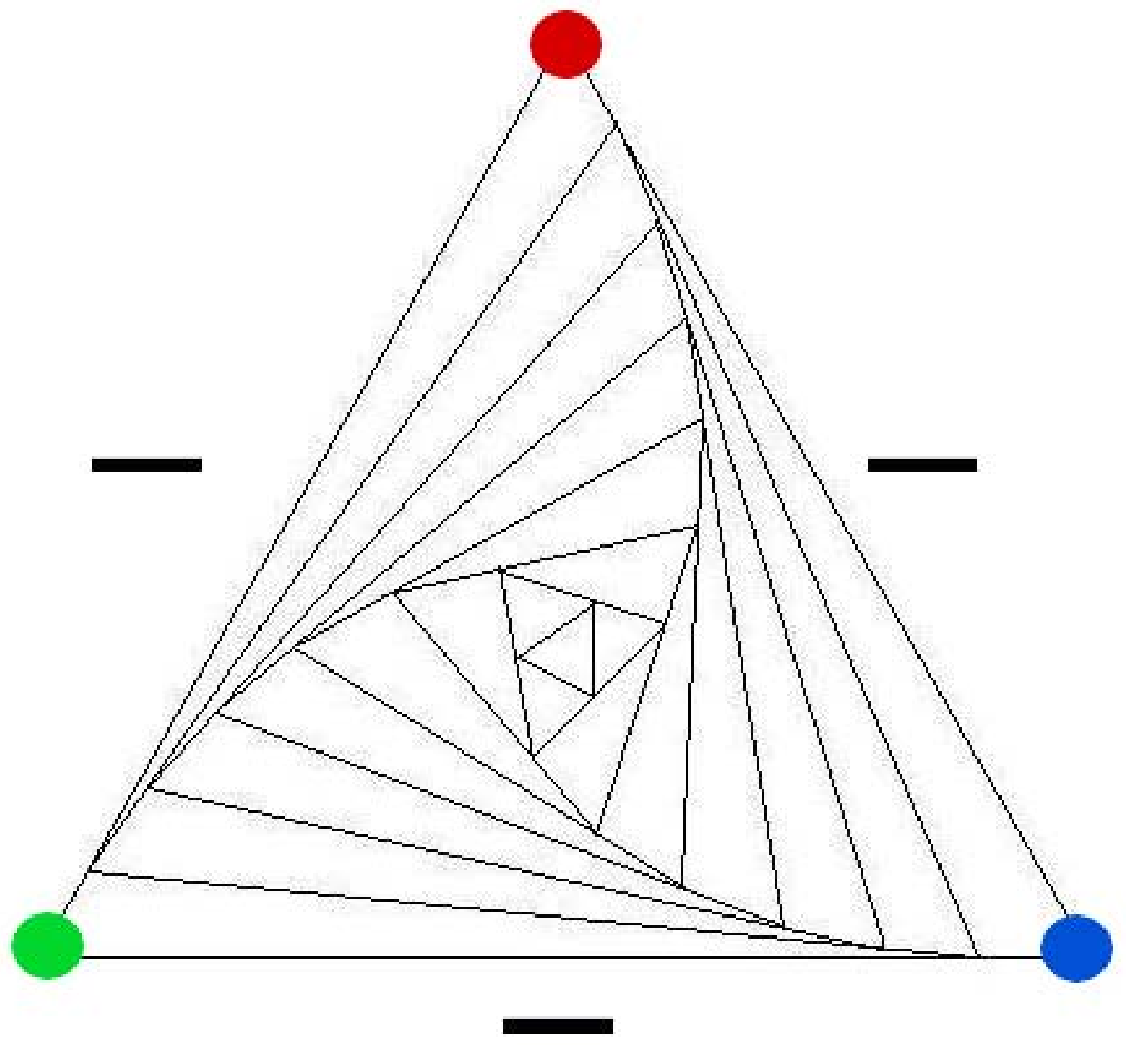} 
 	\vspace{-5mm}
\end{minipage} 

\  

\

\caption{The 3-Vertex cliques (triangles) are efficient indicators of a network structure. In the cognitive networks is frequently possible to characterize the links, for instance as \textit{positive} or \textit{negative}, in this case there are only 4 possible undirected triangles (e.g. X-Full Positive (+++);Y-Hybrid Positive (+-+); $\Omega$ -Hybrid Negative (--+); W-Full Negative (---)).}\label{Triangles_ex}
\hspace{10mm}%
\end{figure}

The discretization of the links allowed to compute the number of polarized 3-vertex cliques, or triangles. The triangles, where each vertex represents an individual, are the 3-cliques of the network created by means of equation~(\ref{pippo}). The four  possible triangles are represented in Fig. \ref{Triangles_ex}, and have been labelled  as $X$,$Y$,$\Omega$ and $W$-triangles. 

For each experiment we compared the number of triangles of each type with the average number of the polarized triangles in $100$ networks containing $10$ nodes and generated by randomly assigning the same number of positive and negative links, and eventually computing  a \textit{z}-score for each of triangle types, a dimensionless quantity that indicates how many standard deviations an observation or datum is above or below the mean. As usual, we adopted as a threshold the value of $\textit{z}=1.96$ that indicates a result corresponding to a marginal probability of $5\%$.

\section{Results}
\label{Section: Results}

\begin{figure}[h]
  \begin{center}
    \begin{tabular}{cc}
    \includegraphics[width=0.45\columnwidth]{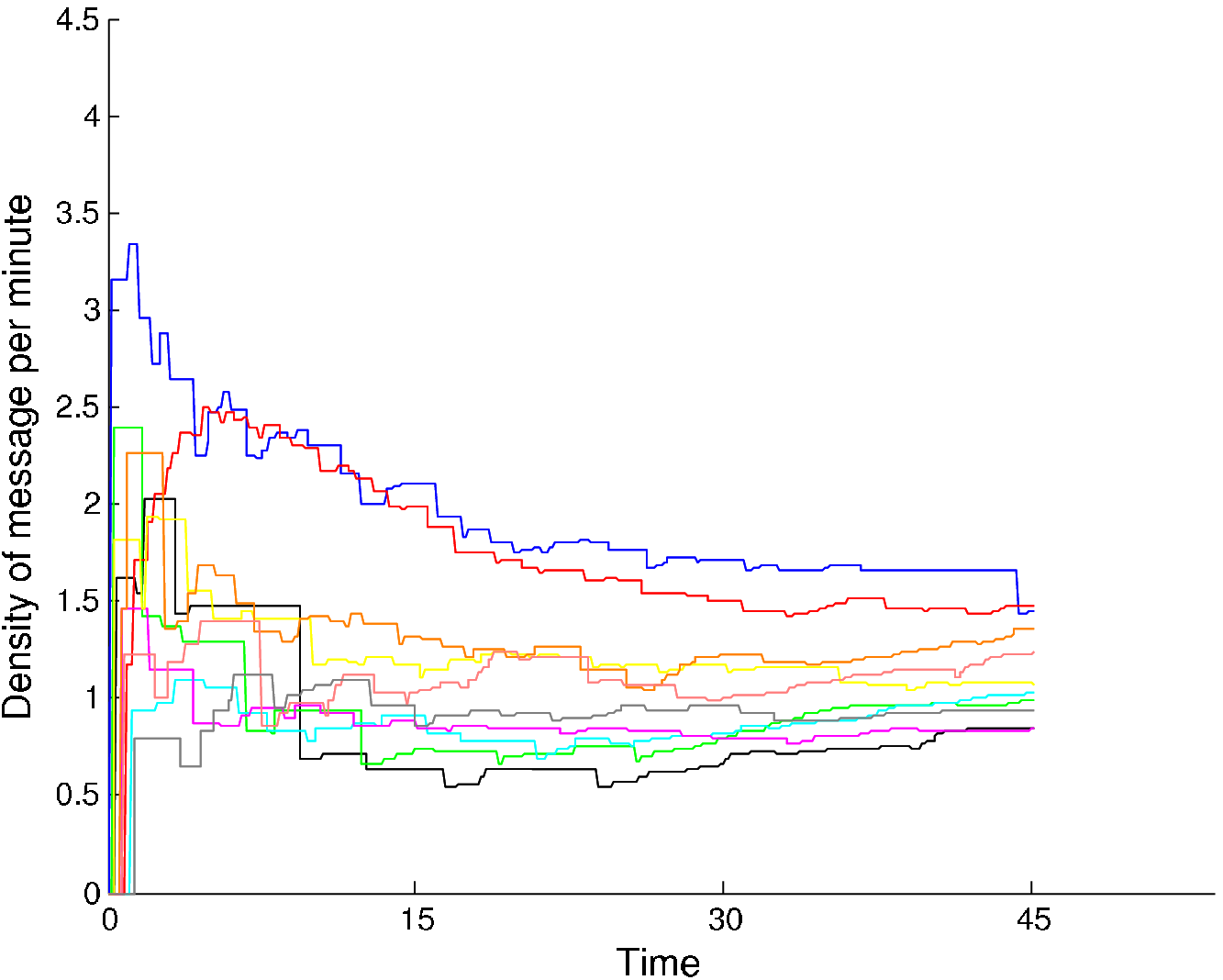}&
    \includegraphics[width=0.45\columnwidth]{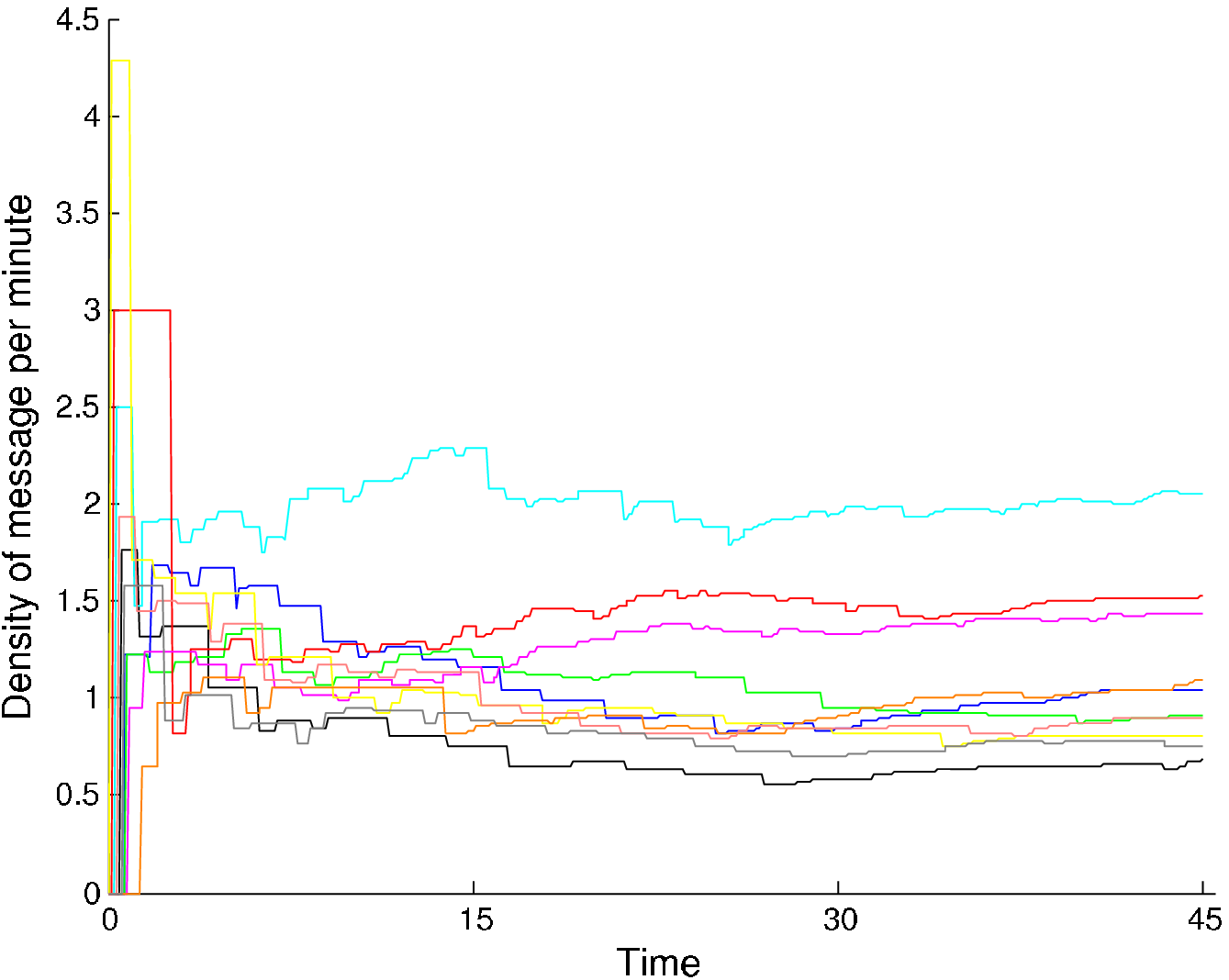}\\
    \end{tabular}
  \end{center}
    \caption{Time evolution of the average agents' normalized activity in the public space (average number of messages per minute), for both Blank modality (left) and Topic modality (Right)}  \label{fig:activities_Public} 
\end{figure}

All the experiments look very similar in terms of node activity for the public environment, while they show a great variability in the private messages space. We report two generic examples in Fig. \ref{fig:activities_Public},  for the Blank and the Topic modality. The public activity of all nodes seem to converge towards a typical value and appear to become quite stable after the first half of the experiment. This behavior appears to be the same for all the $10$ experiments. The typical value of the average number of messages variates more  among experiments than between the two experimental modalities. 

The weighted centrality of the $10$ subjects within this space appears to converge for all the experiments to the critical value of $\frac{1}{N(N-1)}$, \textit{ i.e.}, the value of a fully connected network.

\begin{figure}[h] 
  \begin{center}
    \begin{tabular}{cc}
    \includegraphics[width=0.45\columnwidth]{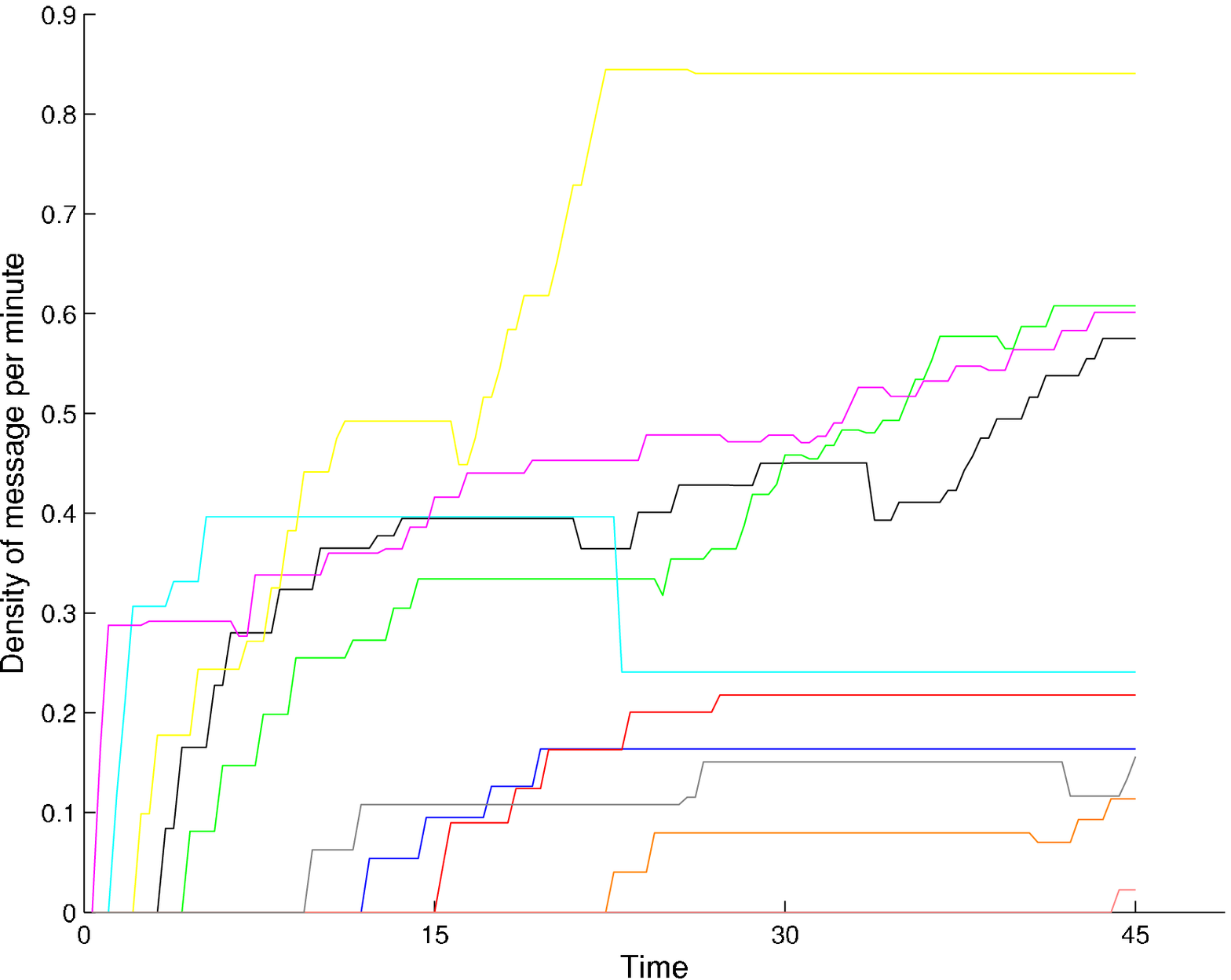}&
    \includegraphics[width=0.45\columnwidth]{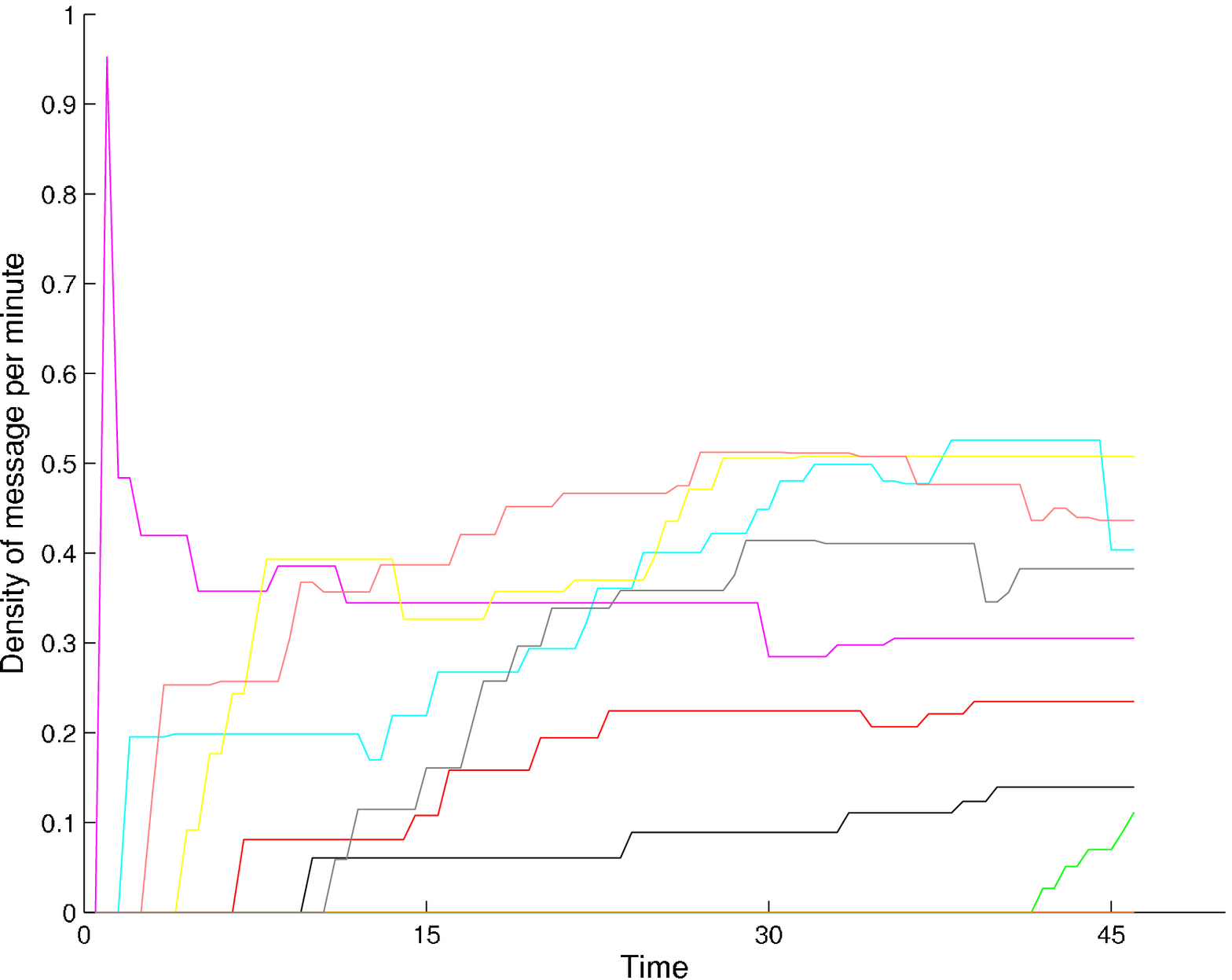}\\
    \end{tabular}
  \end{center}
    \caption{Time evolution of the agents' normalized activity in the private space (average number of messages per minute), for both Blank modality (left) and Topic modality (right)}\label{fig:privato}
\end{figure}

On the contrary, the participant activity in the private space shows a high variability, as reported in Fig.~\ref{fig:privato}.

We analyzed the quantitative differences between the Blank and Topic communicative behavior of the group by means of the ANOVA. We have reported in Table \ref{table:ANOVA}  the communicative variables that are significantly different between the two experimental conditions.
The Blank unconstrained modality is characterized by a larger number of messages than the Topic modality. This quantitative trend describes also the number of messages directed to a single agent as well as the total number of positive messages. 

Interestingly, only the positive mood seems to be able to discriminate between the two experiments, while the negative messages do not appear to differ significantly.

Concerning the private radar management (\textit{i.e.}, the affinity assessment process), the average number of displacements on the radar has been larger for the topic modality as well as the average normalized distance among the icons.    
\begin{center}{
\begin{table}[ht]
\caption{Analisys of Variance between \textit{Blank} and \textit{Topic} related variables} 
 \centering 
\begin{tabular}{crrr} 
\hline\hline 
\textbf{Communicative Variable} & \multicolumn{1}{c}{\textbf{F}} & \multicolumn{1}{c}{\textbf{Blank}} & \multicolumn{1}{c}{\textbf{Topic}} \\ [0.5ex] 
\hline 
Messages & 8.54* & 310 & 228\\ 
\textit{Directed} messages & 23.21* & 38 & 21\\
Positive messages  & 34.22* & 23& 8\\
\hline
Public Messages & 7.31* & 296 & 222 \\
Public Positive Messages & 21.28* & 203 & 96 \\
Private Messages & 8.77* & 14 & 6 \\
Private Positive Messages & 19.31* & 6 & 1 \\
\hline
Displacements: Private Radar & 7.92* & 1.4 & 2.4 \\
Average Distance: Private Radar & 11.81* & .098 & .123 \\[1ex] 
\hline 
\end{tabular}
\label{table:ANOVA}\\ 
* : $\textit{p.} < .01$
\end{table}
}\end{center}

We investigate the differences in  private radar behavior with respect to the experimental modality by means of the Student \textit{t} analysis, Table \ref{table:StudentT}.
We considered the two groups defined by the Eq. \ref{eq:Affinity_recoding}  comparing the typical behaviors towards a \textit{close} agent with that towards a \textit{far} one, for both the Blank and the Topic cases.
The  typical strategy for the Blank modality seems to be quite evident using the communicative variables we have registered. Agents exchange more messages with their inner circle than with the outer one. The variables which appear to be significantly affected by the task are the total number of positive messages, and the number of public and private total and positive messages.

Remarkably, no significant difference emerges for the topic modality, where the strategy assessment of affinity seems to be unrelated with the communicative dynamics.


\begin{center}{
\begin{table}[h]
\caption{Student \textit{t} on communicative variables between \textit{Close} (A) and \textit{Far} (B) agents on the private radar } 
 \centering 
\begin{tabular}{|c|r|r|r|} 
\hline\hline 
\multicolumn{4}{c}{\textbf{Blank modality}}\\ 
\hline\hline 
\textbf{Variable} & \multicolumn{1}{c|}{\textbf{t}}& \multicolumn{1}{c|}{\textbf{A}} & \multicolumn{1}{c|}{\textbf{B}}\\ 
\hline 
Average Radar Distance & \multicolumn{1}{c|}{-} & 0.31 & 0.82\\  
\hline 
Global Messages & 4.84* &  0.15 & 0.10 \\
\hline 
Positive Global Messages & 3.22* &  0.15 & 0.09 \\
\hline 
Public Messages & 3.53* &  0.13 & 0.10 \\
\hline 
Public Positive Messages & 3.26* &  0.12 & 0.10 \\
\hline 
Private Messages & 3.36* & 0.15 & 0.05 \\
\hline 
Private Positive Messages & 3.69* &  0.09 & 0.01 \\
\hline\hline 
\multicolumn{4}{c}{\textbf{Topic modality}}\\
\hline\hline 
\multicolumn{4}{|c|}{No significant difference}\\ 
\hline 
\end{tabular}
\label{table:StudentT}\\ 
* : $\textit{p.} < .01$
\end{table}
}\end{center}


\begin{figure}[h]
  \begin{center}
    \begin{tabular}{cc}
    \includegraphics[width=0.45\columnwidth]{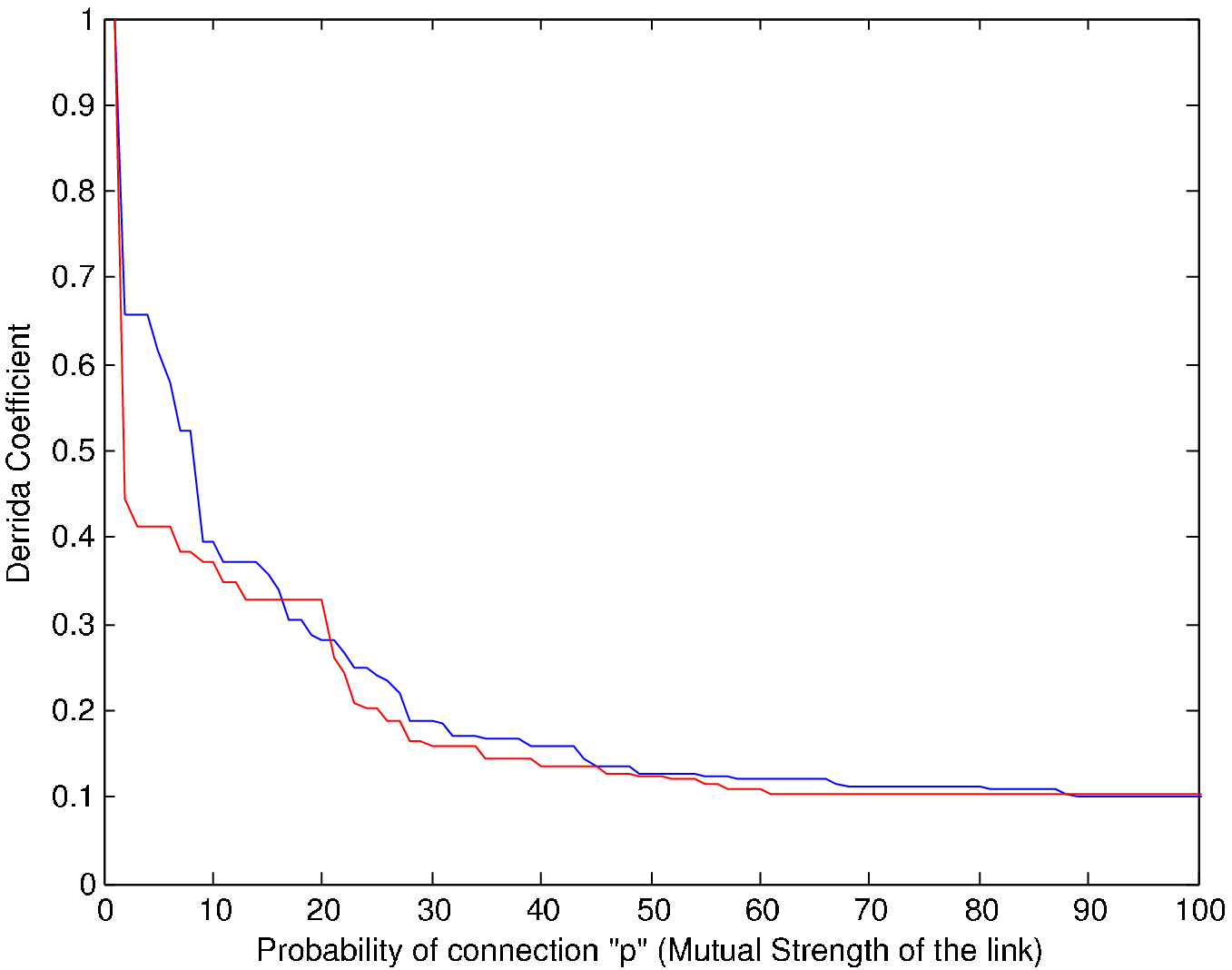}&
    \includegraphics[width=0.45\columnwidth]{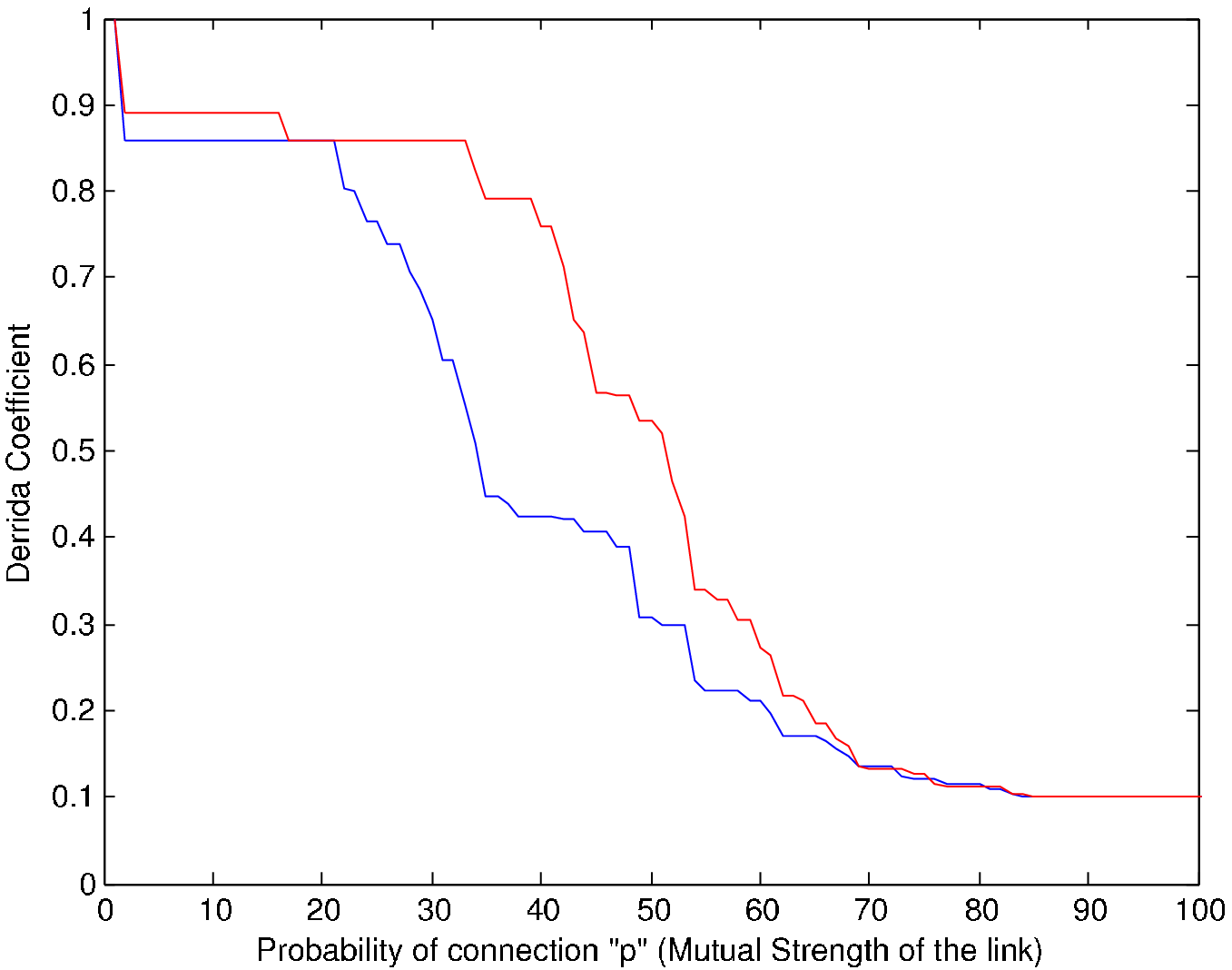}\\
    \end{tabular}
  \end{center}
    \caption{Derrida spectrum for Blank modality (blue line) and Topic modality (red line) is reported for the private messages space (left plot) and for the private radar space (right plot)}\label{fig:Derrida} 
\end{figure}


\begin{center}{
\begin{table}[h]
\caption{ Frequencies of $\Omega$-triangles and \textit{z}-scores for Blank and Topic modalities} 
 \centering 
\begin{tabular}{|c|r|r|r|r|r|r|r|r|} 
\hline\hline 
\textbf{} & \multicolumn{4}{c|}{\textbf{Blank}} & \multicolumn{4}{c|}{\textbf{Topic}}\\ 
\hline\hline
\textbf{Exp} & \multicolumn{1}{c|}{\textbf{+}} & \multicolumn{1}{c|}{\textbf{-}} & \multicolumn{1}{c|}{\textbf{$\Omega$}} & \multicolumn{1}{c|}{\textbf{\textit{z}$\Omega$}} & \multicolumn{1}{c|}{\textbf{+}} & \multicolumn{1}{c|}{\textbf{-}} & \multicolumn{1}{c|}{\textbf{$\Omega$}} & \multicolumn{1}{c|}{\textbf{\textit{z}$\Omega$}}\\ 
\hline 
1 & 18 &  34 & 14 & 1.4 & 26 & 38 & 23 & 1.3\\
\hline 
2 & 22 &  40 & 18 & 0.1 & 18 &  50 & 29 & 1.8\\
\hline 
3 & 16 &  28 & 9 & 1.4 & 14 &  48 & 24 & 2.6\\
\hline 
4 & 10 &  48 & 16 & 1.7 & 32 &  40 & 33 & 2.1\\
\hline 
5 & 12 &  30 & 3  & -1.2 & 14 &  54 & 32 & 3.6\\
\hline\hline 
Average Value & 15.6 & 35.9 & 12 & 0.68 & 20.8 &  46.1 & 28.2 & 2.28\\
\hline\hline 
\end{tabular}
\label{table:Triangle_Omega}\\ 
\end{table}
}\end{center}


In Fig. \ref{fig:Derrida} we show the average Derrida coefficient functions for the private messages space and for the private radar space, for the Blank and the Topic modalities. As we will discuss in the next section, by means of the Derrida coefficient we investigate the typical dispersion of the icons. The two figures show some differences in the distribution of the cluster size with respect to the experimental conditions.

We finally investigated the density of the \textit{polarized triangles} (see Section \ref{Statistical and Structural Analysis}).

First for each experiment we computed the number of positive and negative links among subjects. We also computed the number of the four possible polarized triangles. The data for the Y triangles are reported in Table \ref{Triangles_ex} for all the experiments.

\section{Discussion and Conclusions}
\label{Section: Discussion}

The goal of our experiment was to discover the strategies (heuristics) that the subjects adopted to cope with the social problem posed by the tasks: assess own affinity with others in presence and absence of an imposed topic in a virtual chatline, without considering the semantic content of messages. We assume that the provided environment is not affecting the emotive/cognitive processes so that our results are valid also for a second task we furthermore asked the participants to reach consensus at least in some groups.

We assumed that the proposed tasks are so common in everyday life that all subjects owned well-established personal strategies.
We  tried to put into evidence the similarities and differences in strategies between the two tasks.
We focused the analysis on structural data (number of messages, timing, position of graphical elements), trying to focus on context-free characteristics of the group dynamics.

A common ingredient of many recent studies in the field of socio- and econo-physics 
is the concept of affinity (or social distance) among subjects. The affinity is considered a fundamental parameter to model the opinion evolution of the group. In our experiment we explicitly asked participants to assess their affinity or distance with others by means of icons on a display called private radar. Consequently, we investigated the relationship between the communicative dynamics and the personal assessment of affinity, as revealed by the private radar.

First of all, we can notice that the public activity of subjects in the two modalities seems approximate the same regime, (Fig.~\ref{fig:activities_Public}). A common feature of the average activity and centrality is that
in all the experiment the relative rank of subject shows little variation (the more prolix remains so, similarly to the less prolix). However,
 after the half of the experiment period,  all subjects' activity becomes rather stationary.

We can interpret this behavior by assuming that the average time needed to establish a group structure (speaker hierarchy, leader role, etc.) is about 20 minutes, for a group of 10 participants. This implies that we can safely perform measurements on the second half of the experimental period and consider them as "stationary" quantities

On the contrary, the participant behavior in private is quite different and is much more changeful (Fig.~\ref{fig:privato}). This seems to imply that the formation of allies and cliques is a much slower process, probably due to the complexity of mutual relationships. Therefore, we can expect that the private communications are much more informative of the group dynamics than the public ones.

The analysis of variance of several quantities in the two modalities (Table \ref{table:ANOVA}) reveals other informations about the differences in communication strategy and affinity assessment. The Topic modality, having a quite stricter requirement, is characterized by a smaller number of messages but larger displacements on the private radar.
Also the normalized average distance in the private radar is larger in the Topic modality. Since the normalization puts the nearest icon at the origin, and the farther at normalized distance one, a larger average distance indicated a more skewed distribution of icons.

The Blank modality distribution (blue line) suggests that no typical structures emerge from the private dynamics, and consequently a smooth decreasing distribution represents the relation between the cluster size and their probability of existence.

In Topic distribution (red line) a typical coarse graining of the network clusters seems to emerge. In spite of the noise in data,    a plateau is detectable for the Derrida value of $0.32$, suggesting  a partition in three groups for the final communicative private network.
For large $p$, the two curves became almost undistinguishable indicating that the short distance distribution is independent from the task.

The same analysis has been conducted on the private radar space, where by means of the Derrida coefficient we investigate the typical dispersion of the icons. The average Derrida coefficient functions for the private messages space  and for the private radar space are reported in Fig. \ref{fig:Derrida}, for the Blank and the Topic modalities.
Since the Derrida coefficient is an indicator of the number of clusters, going from 1 (all in a single cluster) to $1/N$ ($N$=number of nodes) for no clusters at all~\cite{Wuensche:2011}.
Both curves have a similar sigmoid shape, and the same asymptotic behavior, however for small $p$ the Blank modality has a longer initial plateau. Nevertheless, in the Topic modality (red line), the icons appear characterized by a larger average distance from the centre but less dispersed than in the blank modality.

Our interpretation of the results is that people in the Blank modality is more concentrated in showing themselves, while in a topic modality one gives more attention in reading other messages and trying to infer other's mutual relationship.
On the other hand, the number of polarized links is larger in the Topic modality than in the Blank one.

In particular, whilst the X, Y and W triangle distributions are not significantly different from the null hypothesis in all the experiments and both modalities,
we found that the $+--$ ($\Omega$) triangle distribution in the Topic modality (Table \ref{table:Triangle_Omega}) is significantly different from the null-hypothesis, revealing a strategy that goes beyond the pair relationships. Indeed, such a clique could represent a situation in which two individuals are allied against a third one, suggesting among the possible interpretations, that the cognitive strategy adopted by the subjects is inspired by the principle \lq\lq the enemy of my enemy is my friend\rq\rq. 

According to such a strategy, only the cliques X and $\Omega$ are stable, while the Y and W are unstable. So, if the inner dynamics of the system is pairwise, not only a final situation with all negative links is also possible, but the density of $\Omega$ triangles should be the same as in a system where the sign of the links is uniformly picked at random. Instead, in the topic modality part we see that the final distribution is not consistent with a pairwise interaction among individuals, meaning that every agent sets his/her relationship with another one taking care of his/her entire neighborhood, and \lq\lq the enemy of my enemy is my friend\rq\rq can be assumed as a possible good strategy to form alliances against a common opponent, and reach a stable configuration. 

The angular distribution of icons in the private radar (both modalities) however does not show any tendency to cluster: people used this instrument to remember others' distance, but not for remembering alliances. The analysis of the polarized 3-vertex cliques could be a precious observable for the human network analysis \cite{Szell2011,Leskovec2010}.

The analysis of the Student \textit{t} confirms the use of different strategies of affinity assessment in the two modalities (Table \ref{table:StudentT}). In the Blank modalities, there is a large correlation among the exchanged messages and the final affinity, while this is not true for the Topic modality. 

While participants in the Blank modality seem to follow the classical assumptions of opinion formation theories, where the frequency of interactions is determinant for diminishing the social distance (or increasing affinity) and consequently determining the convergence of opinions, this behavior is not present in the Topic modality, although part of this task concerned the formation of group of coherent opinion. 

We interpret this difference as a signature of the adoption of different strategies or heuristics. We suppose that in the Topic modality the semantic content of exchanged messages played a major role, and that participants preferred to infer other's opinion by observation and not by direct communication. 

This supports the assumption by Lewin about a strong group effect on individual behavior (psychological field); most of quantities that we analyzed do not exhibit differences at the level of the single individual in the two modalities, except the explicit assessment of affinity with others, which is a group-related quantity. From these results it seems that, at least in a controlled chat experiment, it is not possible to understand (and modeling) the group behavior starting from the observation of a single individual, and on the other hand it is not possible to understand the evolution of the individual cognitive heuristics without taking into account the interdependency between the individual strategies and the group structure.
Further investigations and an analysis of the semantic content of messages are required in order to confirm these hypotheses and extend them in real-life situations.

\section*{Acknowledgments}
This work was partially funded by the European Commission under the FET-AWARENESS RECOGNITION (257756).



\begin{thebibliography}{50}
\bibitem{Cattell1966} Cattell, R. B. . \emph{The scree test for the number of factors}. Multivariate Behavioral Research,  \textbf{1}, Pages 245–276, (1966)\\
\bibitem{Comrey1992} Comrey, A., \& Lee, H. . \emph{ A first course in factor analysis}. Hillsdale, NJ, England: Lawrence Erlbaum, (1992).
\bibitem{Newman} M.E.J. Newman and M. Girvan, \emph{Finding and evaluating community structure in networks}, Phys. Rev. E \textbf{69}, 026113 (2004). 
\bibitem{Fortunato} S. Fortunato and C. Castellano, \emph{Community structure in graphs}, Springer's Encyclopedia of Complexity and System Science  (Springer, Berlin, 2010) pp. 1141--1162.
\bibitem{Zhang1994} J. Zhang, D.A. Norman, \emph{Representations in Distributed Cognitive Tasks}, Cognitive Science, \textbf{18}, 87-122 (1994).
\bibitem{Lewin1943} K., Lewin \emph{Defining the Field at a Given Time}, Psychological Review, \textbf{50}, 292-310 (1943). 
\bibitem{Leavitt1951}  H.J., Leavitt, \emph{Some effects of certain communication patterns on group performance.},Journal of Abnormal and Social Psychology, \textbf{46}, 38–50 (1951).
\bibitem{Bavelas1950} A., Bavelas, \emph{Communication patterns in task-oriented groups}, Journal of the Accoustical Society of America, \textbf{22}, 725–730 (1950).
\bibitem{Bion} W. R., Bion, \emph{Group dynamics: a review},  International Journal of Psycho-Analysis, \textbf{33}, (1952); 
\bibitem{Vilone2011} D. Vilone, A. Guazzini, \emph{Social Aggregation as a Cooperative Game}, Physica A, \textbf{390}, 2716 (2011).
\bibitem{Gigerenzer1996} G. Gigerenzer, G. Goldstein, \emph{Reasoning the Fast and Frugal Way: Models of Bounded Rationality}, Psychological Review, Vol. 103, \textbf{4}, 650-669 (1996). 
\bibitem{Gigerenzer2011} G. Gigerenzer, G. Goldstein, \emph{The recognition heuristic: A decade of research},Judgment and Decision Making, Vol. 6, \textbf{1}, 100–121 (2011).
\bibitem{Simon1976} H. A. Simon, J. R. Hayes, \emph{The understanding process: Problem isomorphs}, Cognitive Psychology, \textbf{8}, 165-190 (1976).
\bibitem{Neisser1967} U. Neisser, \emph{Cognitive psychology}, Appleton-Century-Crofts New York (1967).
\bibitem{Festinger1950} L. Festinger, S. Schachter, K. Back, \emph{Social Pressures in Informal Groups: a Study of Human Factors in Housing.}, Palo Alto, California: Stanford University Press, 1950.
\bibitem{Guazzini2010} A., Guazzini, P., Li\`o, F., Bagnoli, A., Passarella, M., Conti, \emph{Cognitive network dynamics in chatlines}, Procedia Computer Science, Volume 1, Issue 1, May 2010, Pages 2349-2356. (2010)\\
\bibitem{Szell2011} M. Szell, R. Lambiotte and S. Thurner, \emph{Multi-relational Organization of Large-scale Social Networks in an Online World}, PNAS  \textbf{107}, 13636 (2010).
\bibitem{Brautbar2011} M. Brautbar, M. Kearns \emph{A Clustering Coefficient Network Formation Game}, Symposium on Algorithmic Game Theory (SAGT)  \textbf{arXiv:1010.1561v2}, (2011).
\bibitem{Nguyen2009} V Nguyen, Z Koukolikova, F Bagnoli, P Lio’, \emph{Noise and nonlinearities in high-throughput data}, J. Statistical Mechanics, P01014, (2009).
\bibitem{Crowne} D. P., Crowne, D., Marlowe, \emph{ A new scale of social desirability independent of psychopathology}, Journal of Consulting Psychology, \textbf{24}, Pages 349-354, (1960)\\
\bibitem{Wuensche:2011} Wuensche, A. \emph{Exploring Discrete Dynamics}, chapter 22,  Luniver Press (2011).
\bibitem{Leskovec2010} Leskovec J, Huttenlocher D, Kleinberg J., \emph{Predicting positive and negative links in online social networks}, ACM WWW Int Conf on World Wide Web. (2010).
\end{thebibliography}
\end{document}